\documentclass[titlepage]{csetr}
\usepackage{times}
\usepackage{subfigure}
\usepackage[pdftex]{graphicx}
\usepackage{amssymb}
\usepackage{caption}
\usepackage[ruled]{algorithm2e}
\usepackage{array}

\renewcommand{\and}{\hspace{.5cm}}

\trnumber{201325}
\trdate{October 2013}

\title{%
 Scalable Protein Sequence Similarity Search using Locality-Sensitive Hashing and MapReduce 
}
\author{%
  Freddie Sunarso$^1$ \and %
  Srikumar Venugopal$^1$ \and\\ %
  Federico Lauro$^2$\\[2em]
  $^1\, $School of Computer Science and Engineering,\\%
  University of New South Wales, Sydney, Australia \\%
  \email{\{freddies,srikumarv\}@cse.unsw.edu.au}\\
  $^2\,$School of Biotechnology and Biomolecular Science, \\%
  University of New South Wales, Sydney, Australia \\
  and \\%
  Singapore Centre on Environmental Life Sciences Engineering,\\%
  Nanyang Technological University, Singapore\\%
  \email{flauro@unsw.edu.au}\\[3cm]
}

\date{}
\begin{document}
\maketitle

\begin{abstract}
Metagenomics is the study of environments through genetic sampling of their microbiota. Metagenomic studies produce large datasets that are estimated to grow at a faster rate than the available computational capacity. A key step in the study of metagenome data is sequence similarity searching which is computationally intensive over large datasets. Tools such as BLAST require large dedicated computing infrastructure to perform such analysis and may not be available to every researcher. 

In this paper, we propose a novel approach called ScalLoPS that performs searching on protein sequence datasets using LSH (Locality-Sensitive Hashing) that is implemented using the MapReduce distributed framework. ScalLoPS is designed to scale across computing resources sourced from cloud computing providers. We present the design and implementation of ScalLoPS followed by evaluation with datasets derived from both traditional as well as metagenomic studies. Our experiments show that with this method approximates the quality of BLAST results while improving the scalability of protein sequence search.
\end{abstract}

\section{Introduction}
\label{sec:Intro}

Metagenomics, which is the study of uncultured microorganisms from their habitats, is an expanding field~\cite{Metagenomics}. A typical metagenomics workflow starts with taking genetic samples and metadata from a particular environment. The next step is to extract DNA from this sample, followed by library construction, sequencing, read preprocessing, and assembly~\cite{Metagenomics2}. The resulting genomic sequences are then processed in order to perform gene calling (prediction) and annotation as well as classification. In this step, sequence similarity searching is performed to identify genes that match with known homologs. The output of sequencing machines from the previous step is compared to the current known databases in order to obtain meaningful information from the sequences. Tools such as BLAST (Basic Local Alignment Search Tool)~\cite{BLAST,GappedBLAST} are widely employed for this purpose~\cite{wooley_metagenomics_2010}. 


In recent years, so-called Next-Generation Sequencing Machines have boosted the speed at which genomes can be sequenced from environmental samples~\cite{shendure_next-generation_2008,loman_high-throughput_2012}. This in turn has led to a deluge in the amount of available metagenome data that is estimated to double every 14 months~\cite{wooley_metagenomics_2010}. Similarity searching is a key bottleneck in the workflow with requirements for clusters with hundreds or thousands of computational cores to execute BLAST on terabytes of data~\cite{CAMERA}. 

A surfeit of tools and techniques such as MPI-BLAST~\cite{MPIBLAST,lin_coordinating_2011}, ScalaBLAST~\cite{oehmen_scalablast_2006}, RAPSearch~\cite{RAPSearch,RAPSearch2}, and LAST~\cite{LAST}, among others, have attempted to improve the performance and scalability of sequence similarity searching. However, as we'll discuss in Section~\ref{sec:relworks} and Section~\ref{sec:Evaluation}, most of these use tightly-coupled concurrent processes communicating via shared memory over dedicated computing infrastructure. This limits their applicability to research groups with the ability to access such infrastructure at scale.  

Infrastructure as a Service (IaaS) (or cloud computing) enables users to provision or de-provision computing resources on-demand through a self-serve interface. The users are charged only for the capacity used on an on-going basis (“pay-as-you-go”) with little or no upfront costs. Recently, there have been many new bioinformatics tools and projects~\cite{odriscoll_big_2013,CloudBLAST,wilkening_using_2009,wall_cloud_2010} that have attempted to take advantage of the convenient access to large-scale computing resources provided by IaaS. 

However, elasticity in the infrastructure causes execution and performance issues for scientific applications that are reliant on complex communications using a shared memory model~\cite{ekanayake_high_2010,he_case_2010}. Therefore, it has been suggested that programming for clouds requires novel paradigms and algorithms~\cite{larus_programming_2010}. This motivates the development of new methods to scale protein similarity search over cloud infrastructure.

Similarity searching is essentially a nearest neighbour problem, that is, given a collection of points (objects) in a $n$-dimensional space, and a query point, the algorithm has to be able to return the points that are closest to the query. This is difficult when $n$ is high, as in the case of protein sequences. Locality-Sensitive Hashing (LSH)~\cite{LSH} is a method originally developed to alleviate the difficulties of performing similarity searching in data with a high number of dimensions.  In LSH, the nearest neighbour problem is approximated by hashing the points using several hash functions to ensure that the probability of collision is much higher for objects close to each other and lower for objects located far apart.  LSH has been used primarily for searching through the very large collection of web documents in the internet, such as demonstrated in several well-known search engines~\cite{BroderWeb, Minhash, MankuSimhash}.

In this paper, we propose a novel method to significantly improve the scalability of protein homology searching by utilising LSH. Similarity searching is performed by first generating sketches, or signatures, of each sequence, and then measuring the Hamming distance between the signatures of the query and reference sequences. We call our approach ScalLoPS (\emph{Scal}able \emph{Lo}cality-Sensitive \emph{P}rotein \emph{S}imilarity Search). Our contributions are twofold:
\begin{itemize}
  \item  We adopt and modify a specific implementation of LSH, proven to be able to quickly approximate the cosine distance between two documents~\cite{Simhash}, for the purpose of high performance searching through protein sequences, and
  \item  We implement our method using the MapReduce~\cite{MapReduce} programming paradigm to distribute the workload across multiple nodes in a scale-agnostic fashion.
\end{itemize}

We also carry out experiments to measure the performance, scalability, and the quality of the method by running the process with actual sets of query sequences, obtained from the bacteria \emph{Escherichia coli} and from the Global Ocean Sampling (GOS) project~\cite{GOS}, against several known reference datasets that are widely available.

The rest of the article is structured as follows. Section \ref{sec:relworks} provides related works. Section \ref{sec:LSH} describes the LSH solution proposed in this article with the MapReduce implementation in Section \ref{sec:MapReduce}. Then, Section \ref{sec:Evaluation} evaluates the implementation, and finally Section \ref{sec:FutureWorks} provides conclusion and planned future works.

\section{Background and Related Work}
\label{sec:relworks}

\subsection{Basic Local Alignment Search Tool (BLAST)}

BLAST is the most popular tool for searching for homologies in genomic and proteomic databases. As BLAST is considered as the ``gold standard'' for sequence similarity search, we present a summary of its workflow in Algorithm~\ref{alg:BLAST} and a brief description in the following paragraphs, so as to make the context clearer as well as to contrast against our approach presented in later sections. 

\begin{algorithm}
 \caption{Summary of BLAST}
 \SetAlgoLined
  \KwData{Reference sequences and query sequences}
  \KwResult{A list of aligned pairs of matching reference sequence and query sequence}
  \ForAll{$seqeuence \in querySequences$} {
      \ForAll{$k$-$token \in sequence$}{
		$neighwords \gets$ generateNeighwords($k$-$token$)\\
		\ForAll{$neighword \in neighwords$}{
				$seeds \gets$ FindMatch($neighword$)\\
				$highScoringPairs \gets$ \{\}\\
			\ForAll{$seed \in seeds$}{
				$highScoringPairs \gets$ extend($seed$)
			}
			\ForAll{$highScoringPair \in highScoringPairs$}{
				Evaluate($highScoringPair$)\\
				Align($highScoringPair$)
			}
		}
      }
    }
\label{alg:BLAST}
\end{algorithm}

\subsubsection{Tokenization}
BLAST performs firstly by tokenizing an input query sequence into $k$-letter words. For example, a query sequence is as follows: "WDERKQYT..." With k=3, BLAST will tokenize the sequence into "WDE", "DER", " "ERK", "RKQ", "KQY", "QYT", and so on until the final letter of the sequence is included.

\subsubsection{Neighbouring Words Generation}
For each of the token, a list of neighbouring words is generated by looking up a substitution matrix. One widely used matrix is BLOSUM62~\cite{henikoff_amino_1992}. The matrix provides a substitution score for each of the amino acid letters. For example, the BLOSUM62 score of substituting "WDE" into "ADE" is $ = -3 + 6 +5 = 8$. The score will then be taken into account when generating the list of neighbouring words. Only words which score below a certain threshold $T$ are included in the list. If $T$ is 13, for instance, then "ADE" is included in the list of the neighbouring words of "WDE".

\subsubsection{HSP Generation}
BLAST would then search the whole database for exact matches of all the lists of neighbouring words. Then, each match is used as a seed to extend the match to the right and left directions until either a negative score or the end of the sequence is encountered. The resulting match is called a high-scoring pair (HSP). In the following example, "ERK" is a seed and it is found in the dabatase that "EKK" is an exact neighbouring word match of "ERK". Extending the seed, the resulting HSP then consists of "DERK" from the query sequence and "EEKK" from the database sequence.\\
Query sequence:	W D E R K Q\\
Database sequence:	L E E K K L\\
Score:			-2 2 5 2 5 -2\\
Accumulated Score: $-2+2+5+2+5+-2=10$

To improve performance, the next generation of BLAST, Gapped-BLAST, joins two HSPs of the same sequences within a distance $d$ from each other into one longer HSP which score is calculated through the substitution matrix as well. HSPs with accumulated score less than a threshold would be kept, or discarded otherwise.

\subsubsection{Assessment of Significance}
The expect value (e-value) of each HSP score would then be calculated to find out the number of times that any database sequence would obtain a score $S$ higher than $x$ by probability. The calculation is as follows:
$E \approx 1-e^{-p(S > x) D}$ where $p(S>x) = 1- \exp (e^{-\lambda (x - \mu)})$. The $\mu$ is calculated as $[\log (Km'n')]/\lambda$ where $m' \approx m-(\ln Kmn)/H$ and $n' \approx n-(\ln Kmn)/H$.

The $m$ and $n$ themselves are the lengths of the query and reference sequences respectively, and the values of the constants for non-gapped alignment with BLOSUM62 are $\lambda=0.318$, $K=0.13$, and $H=0.40$.

\subsubsection{Sequence Alignment}
Finally, sequence alignment is then performed on the HSPs by Smith-Waterman method. The original BLAST generates multiple alignments for the same pair without any gap, while only one alignment with gaps is produced for each pair by Gapped-BLAST.

As can be perceived from the algorithm~\ref{alg:BLAST}, BLAST to some extent, still performs one-to-one comparison when searching the database of reference sequences for matches of neighbouring words. This one-to-one comparison would result in an undesirable time complexity, particularly for very high number of sequences. 


Many tools and techniques have been established to improve the performance of BLAST, particularly when dealing with large query sets.  mpiBLAST~\cite{MPIBLAST} uses the Message Passing Interface (MPI) distributed programming paradigm to distribute BLAST workload across multiple nodes. MPI performs by spawning many processes, each with its own address space that is not accessible by the other processes. Thus, an MPI job is split between processes, and data is moved from one address space to another by cooperative operations.  MPI is therefore suitable for applications with high data exchange between tightly-coupled nodes.

In mpiBLAST, the BLAST database is segmented and each segment is then distributed into each of the processing nodes. BLAST is then run in parallel on each of the segments. During the evaluation, mpiBLAST shows superlinear speedup relative to the number of nodes when the database is larger than the core memory of a single node. Later publications~\cite{meng_high-performance_2010, lin_coordinating_2011} have reported on enhancing performance of mpiBLAST through various techniques.

However, the performance of MPI programs, especially those with communication-intensive processes, is reduced  on cloud resources~\cite{jackson_performance_2010}, as the latencies involved in maintaining a shared memory over a loosely-coupled distributed system slow down individual processes. Moreover, traditional MPI programs operate on static, pre-configured sets of nodes and cannot accommodate elastic infrastructures easily.

ScalaBLAST~\cite{oehmen_scalablast_2006} performs linear scaling by using multiple techniques such as storing the reference database in memory locations shared among distributed processors, allowing processors to write to separate files independently and splitting queries among processes organised into groups. ScalaBLAST employs pre-fetching, wherein blocks of sequence are fetched from one location to another where they may be needed, in order to hide the latency of executing a similarity search against a distributed database. 

With respect to metagenomic studies, as the query sets are orders of magnitude larger than the reference databases, distributing the reference database is less likely to provide a better performance than distributing the actual queries. Also, with large numbers of distributed resources on a cloud, the communication costs of providing a shared memory abstraction, as performed by ScalaBLAST, are likely overwhelm efficiency gains of parallelising queries.    


RAPSearch (Reduced Alphabet-based Protein Similarity Search)~\cite{RAPSearch} is another tool that performs similarity searching and local alignment developed to overcome the limitation of BLAST. RAPSearch firstly compresses each residue (amino acid) of both query and reference sets using a reduced amino acid alphabet, that is by clustering similar amino acids together. Then, the seed-and-extend method is applied to identify maximal exact matches (MEM) between the reduced alphabet sequence of a query and the whole database, and then uses the same heuristic extension algorithm as used in BLAST to extend and evaluate each of these seeds. 

RAPSearch was measured to be able to achieve about 20 to 90 times speedup as compared to BLAST with minimal loss of sensitivity. RAPSearch was subsequently refined into RAPSearch2~\cite{RAPSearch2}, which reduces the memory footprint by utilising a collision-free hash table to index a similarity search database, and capable to perform multi-threading in a multicore machine. However, RAPSearch cannot be executed on distributed resources.
%


\subsection{LSH and Bioinformatics}

Among the first applications of LSH in bioinformatics is the \textsc{lsh-all-pairs} algorithm developed by Jeremy Buhler~\cite{SequenceComparisonLSH}, for finding local similarities in multimegabase genomic DNA sequences. The algorithm iterates the following steps. Firstly, a random hash function is chosen by picking $k$ indices from a set of \{1,\ldots,$d$\} where $d$ is the shingle length. Then, for every shingle of every sequence in the collection, a tuple containing the LSH value and the shingle position is stored. After that, the tuples are partitioned into different classes, each class consists of all the tuples of the same LSH value. Finally, for each of the classes, all sequence pairs of $d$-mers whose tuples are in the class are compared and pairs that differ by at most $r$ substitution are stored. The algorithm was measured to be able to find significant similarities with as little as 63\% identity.

Buhler later developed \textsc{lsh-all-pairs-sim} that is able to find similarities with a technique known as score simulation that simulates substitution score matrices \cite{LSH-ALL-PAIRS-SIM}. This method is applicable not just for DNA sequences, but also for protein sequences. This method was further refined by a new distance function on peptides derived from a given score matrix that run eight times faster with sensitivity within 5\% of the original \textsc{lsh-all-pairs-sim} \cite{LSH-ALL-PAIRS-SIM2}. Both \textsc{lsh-all-pairs} and \textsc{lsh-all-pairs-sim} are not designed for distributed systems, and therefore are not scalable.

Another application of LSH in metagenomics is sequence clustering. In this article, Rasheed \emph{et al}~\cite{EfficientMetagenomicsClusteringLSH} proposed an algorithm called \texttt{MC-LSH} that utilizes LSH to cluster metagenomic sequences. This method employs an LSH function that computes the subsequence of the original sequence at $k$ random indices. After the LSH values of all sequences are computed, then the following steps are iterated until all sequences are clustered. Firstly, an unassigned sequence is moved to a new cluster label. Then, each of the other sequences in the list of unassigned sequences is assigned to the same cluster label if its hash value differs by at most $r$ percentage of $k$. At the end of this iteration, all the sequences are guaranteed to be clustered. 

This algorithm is measured to be computationally efficient and accurate compared to other quality clustering algorithms~\cite{EfficientMetagenomicsClusteringLSH}. However, such clustering method seems to be not efficient for unequal length datasets which shortest sequence is very short, since it reduces the LSH to letter-by-letter comparison. In this case, the computational complexity is reduced to an undesirable level of $O(n^2)$.

\subsection{Mapreduce and Bioinformatics}

MapReduce~\cite{MapReduce} is a distributed data processing framework based on a simple abstract programming model that leverages the scalability, fault-tolerance and reliability provided by the underlying distributed filesystem. The most popular open-source implementation of MapReduce is produced by the Apache Hadoop project~\footnote{http://hadoop.apache.org/}, and is packaged with a distributed file system known as HDFS (Hadoop Distributed File System)~\cite{HDFS}.

One application of MapReduce in this area is CloudBurst~\cite{CloudBurst}. In this application, DNA sequence data could be mapped to the target genome and other reference genomes by modelling after the short read-mapping program RMAP, which maps reads to known genomes. The map function of CloudBurst produces $k$-shingle as seeds from both the reads and reference sequences, while the reduce function receives sequences with shared seeds and extends them into end-to-end alignments by counting mismatches, or by applying the Landau-Vishkin $k$-difference algorithm for gapped alignment. With this method, a CloudBurst program running on at least 24 processing nodes is measured to be able to achieve more than 24 times speedup over an RMAP program running on a single core processor. Further experimentations also reveal that CloudBurst scales linearly with the number of processing nodes. CloudBurst, however, is designed for mapping reads to known genome, as has been mentioned. In contrast, ScalLoPS is designed to perform sequence similarity searching than read mappings.

Biodoop~\cite{Biodoop} is another application that utilizes MapReduce for parallelizing BLAST. In Biodoop, BLAST is wrapped into a Python module and run within the mapper function, which takes one reference sequence at a time and produces all the alignments with the given query sequences. Biodoop-BLAST is measured to achieve speedup quite linearly with the number of processing nodes. However, this method is simply the distribution of data than computation, while ScalLoPS is designed to distribute both the data and the computation.

The most recent implementation of sequence comparison in distributed systems is the calculation of Smith-Watermann (SW) matrix for sequence alignment on a federated cloud environment, also employing the MapReduce framework~\cite{BioseqComparisonFederatedClouds}. The performance of this approach over a huge genomic database is measured to be comparable to the one obtained in multicore clusters and to achieve up to 22.5 \% speedup over the best stand-alone Cloud execution. Although currently ScalLoPS is not concerned with sequence alignment, this method may be a suitable candidate to further develop ScalLoPS in the future.

\subsection{Comparison to Other Methods}
ScalLoPS is different from other relevant methods presented in this section since it specifically caters for protein sequences, which are more complicated than DNA sequences. Moreover, ScalLoPS also employs both Locality Sensitive Hashing and MapReduce in order to achieve both high performance and scalability in distributed systems, while several other methods presented in this section are not designed for scalability. Table \ref{tbl:RelatedWorks} shows the comparative view of the methods presented in this section against ScalLoPS.


\begin{table} [!ht]
\centering
\caption{Comparative view of the approaches that perform metagenomics processing
\label{tbl:RelatedWorks}}
\begin{tabular} { |p{0.2\textwidth}| p{0.15\textwidth}|p{0.15\textwidth}|p{0.4\textwidth}|}
\hline
Method & Sequence & Approach & Scalability\\
\hline
 \textsc{lsh-all-pairs} \cite{SequenceComparisonLSH} \& \textsc{lsh-all-pairs-sim} \cite{LSH-ALL-PAIRS-SIM, LSH-ALL-PAIRS-SIM2} & DNA/Protein & LSH & (Not designed for distributed computing)\\
\hline
Biodoop-BLAST \cite{Biodoop} & DNA/Protein & MapReduce & Scalability entering saturated region after a cluster size of 21 nodes\\
\hline
MPI-BLAST \cite{MPIBLAST} & DNA/Protein & MPI & Super-linear speedup relative to the number of nodes in the best case \\
\hline
RAPSearch \cite{RAPSearch} \cite{RAPSearch2} & Protein & Multi-threading & (Not designed for distribution over multiple nodes)\\
\hline
ScalLoPS & Protein & LSH/Map\-Reduce & Inverse-exponential scalability\\
\hline
\end{tabular}
\end{table}

\section{LSH-based Protein Sequence Search}
\label{sec:LSH}

As mentioned in Section~\ref{sec:Intro}, LSH is an efficient algorithm to approximate near neighbour search. An LSH function is closely related to the distance metric that it is approximating.  This approximation is carried out by having a family of hash functions to achieve a desired probability of collision such that similar items will collide and be placed in the same buckets. This would happen only when the probability of a hash function $h$ picked from a family of hash functions $H$ mapped two similar documents $x$ and $y$ into the same bucket is equal to the similarity of both documents that is measured according to a certain distance metric:

\textbf{Pr}$_{h \in H}[h(x)=h(y)]=sim(x,y) $

To ensure this, the family must satisfy the properties that close points with distance measured as at most $d_{1}$ have the probability of being a candidate pair at least $ p_{1}$, and that far away points with distance measured at least $d_{2}$ have the probability of being a candidate pair at most $ p_{2}$. Such family is said to be a $ (d_{1},d_{2},p_{1},p_{2})$-sensitive family, and it is obvious that high $p_{1}$ and low $p_{2}$ are desirable.

In particular, we are interested in a specific implementation of LSH that has been proven to be able to quickly approximate the cosine distance between two documents~\cite{Simhash}. Here, data with $d$ number of dimensions are represented by vectors in $\mathbb{R}^{d}$. The similarity between two vectors is then measured by the angles formed between them. The smaller the angle the closer the two vectors are.

LSH can approximate this angle by using a technique known as random hyperplane rounding~\cite{Simhash}. In this approach, suppose that one of the vectors is $\vec{x}$, and a random hyperplane that intersects the plane formed by $\vec{x}$ is selected, with the hyperplane's normal vector denoted as $\vec{r}$. Then, a corresponding hash function ${h}_{\vec{r}}$ can be defined as follows:

$
{h}_{\vec{r}}(\vec{x}) =
\left\{
	\begin{array}{ll}
		1  & \mbox{if } \vec{r}.\vec{x} \geq 0 \\
		0 & \mbox{if } \vec{r}.\vec{x} < 0
	\end{array}
\right.
$

The dot product of $\vec{r}$ and $\vec{x}$ is geometrically the length of a projection of the vector $\vec{r}$ onto the unit vector of $\vec{x}$ and the sign of the dot product indicates the side of $\vec{x}$ relative to the hyperplane. Here the sign is taken as a binary bit, and by having a family of such hash functions, the position of $\vec{x}$ can be approximated by concatenating all the bits from all the hash functions together. This concatenated bits form a signature of the vector.

Generating the signatures for all the vectors in the dataset space, the cosine similarity between two vectors can be measured by the number of bits that differ between them (Hamming distance). This can be formulized as in the following:
\textbf{Pr}$_{h \in H}[h(\vec{x})=h(\vec{y})]=1-\frac{\theta(\vec{x},\vec{y})}{\pi}$.

Therefore, in this method, LSH is carried out by approximating all the angles between two nearest neighbours in $\mathbb{R}$. The similarity of two signatures would then be measured by Hamming Distance, which is the number of the differing bits of the signatures. A low Hamming Distance between the two signatures indicates a high degree of similarity of the sequences and vice versa.

Following this description, Manku \emph{et al} devised a practical way in which this method can be carried out \cite{MankuSimhash}, as elaborated in the following. First, a vector $V$ of $f$ dimensions is initialised to zero. Then, each feature of the input data is hashed into an $f$-bit hash value. If the $i$-th bit of the resulting hash value is 1, then the $i$-th component of $V$ is incremented by the weight of the feature, otherwise, if it is 0, then the component is decremented. After all the features have been taken into account, each of the components of $V$ is either positive or negative. These signs then determine the bits of the final fingerprint.

\subsection{Modification of LSH}

\begin{figure}[t]
\centering
       {\includegraphics[width=0.7\textwidth]{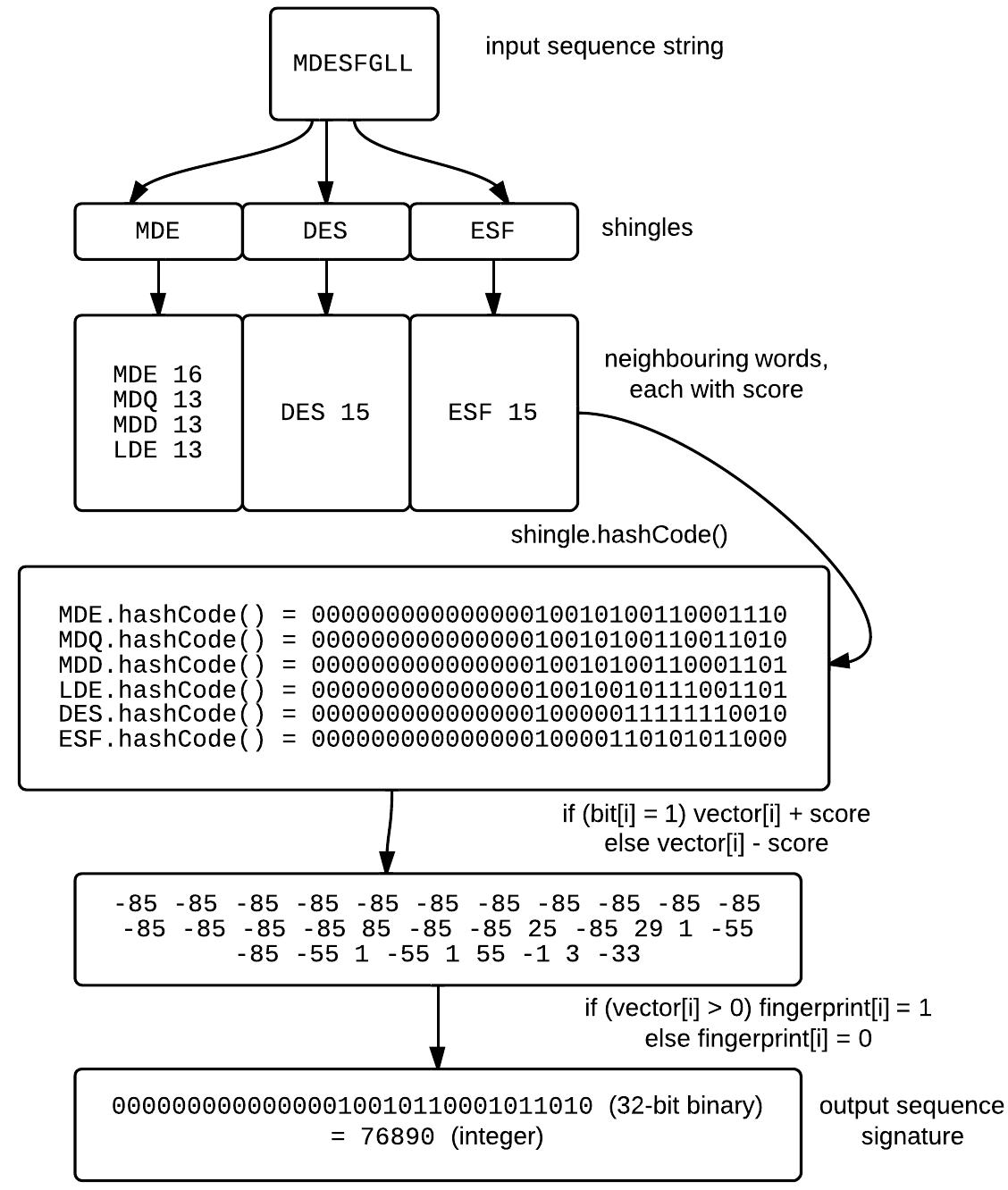}}
 \caption{Example of signature generation with the modified LSH}
 \label{fig:SimhashSigGen}
\end{figure}

As mentioned earlier, LSH is often applied to find similarities in a collection of documents, each of which can be represented by a vector of feature weights. Since a document consists of many words or terms, for such collection a feature is often a word, or a term, of the documents with its frequency being its weight.

In ScalLoPS, however, a string of protein sequence constitutes of its smaller non-overlapping substrings of length $k$, denoted as $k$-shingles, and as such the sequence can be represented by a vector of $k$-shingle substrings. This is similar with the Tokenization step performed by BLAST, as discussed in section \ref{sec:relworks}. For example, as shown in Figure~\ref{fig:SimhashSigGen}, the feature vector of a protein sequence string \texttt{MDESFGLL} is \texttt{\{MDE\},\{DES\},\{ESF\},\{SFG\},\{FGL\},\{GLL\}}.

Then, as shown in Algorithm \ref{alg:SimhashSigGen}, for each shingle substring, a list of neighbouring words can be generated by utilising the BLOSUM62 substitution matrix. In the example of Figure~\ref{fig:SimhashSigGen}, the first shingle \texttt{\{MDE\}} produces neighbouring words such as the \texttt{\{MDE\}} itself along with \texttt{\{MDQ\},\{MDD\},\{LDE\}} by applying BLOSUM62 matrix.

Each of this neighbouring words can then be taken as a feature. Since the importance of each neighbouring word is indicated by its score, the neighbouring word score can then be considered as the feature weight. In the example of Figure~\ref{fig:SimhashSigGen}, the aforementioned feature vector \texttt{\{MDE\},\{MDQ\},\{MDD\},\{LDE\},\{DES\},\{ESF\}} produces the feature weight vector \texttt{\{16,13,13,13,15,15\}}.

Each of the neighbouring words is then hashed with the built-in Java hashCode function for String - 
$hashCode(s)=\displaystyle \sum\nolimits_{i=0}^k s[i].31^{k-i-1}$, $s$ in this case being each of the neighbouring words of the $k$-shingles.

This hash function is selected since it is designed to efficiently minimise the hash value collision for different strings, and will in turn result in almost true random hyperplanes being selected. In the example of Figure~\ref{fig:SimhashSigGen}, the random hyperplanes starting from bit position 16 is $\{1,1,1,1,1\}$, $\{0,0,0,0,0\}$, $\{0,0,0,0,0\}$, $\{1,1,1,1,0,0\}$, $\{0,0,0,0,0\}$, $\{1,1,1,0,0,1\}$, $\{0,0,0,1,1,1\}$, $\{0,0,0,0,1,0\}$, $\{1,1,1,1,1\}$, and so on.

With this modification, each dot product is then carried out between each of the random hyperplane vectors and the weight vector of neighbouring word scores. Figure~\ref{fig:SimhashSigGen} shows an example of such a dot product. Following up the modification above, ScalLoPS generates the signatures of protein sequences by looking at the resulting vector that contains all the dot products, as in the original application of LSH. If the $i$-th component of the vector is equal to or greater than 0, then the $i$-th signature bit is set, otherwise the $i$-th signature bit is not set. In the example of Figure~\ref{fig:SimhashSigGen}, the resulting signature in 32-bit binary is 00000000000000010010110001011010 or equivalent to 76890 in decimal integer.


\begin{algorithm} [t]
 	\caption{Signature Generator Map Function}
	\label{alg:SimhashSigGen}
	\SetAlgoLined
 	\KwData{KEY($sequenceID$), VALUE($sequenceString$)}
	$neighwords \gets$ \{\}\\
	$vector$[0 ... 31] = $\emptyset$\\
	$fingerprint$[0 ... 31] = false\\
	\ForAll{$k$\--$shingles \in sequenceString$}{
		$neighwords = neighwords$ $\cup$  generateNeighwords($k$-$shingles$)\\
		\ForAll{$neighword \in neighwords$} {
			$hashed = neighword.hashCode()$\\
			\If{$hashed[i]$ = true} {
				$vector[i] += neighword.score$
			}
			\Else{
				$vector[i] -= neighword.score$
			}
		}
	}
	\For{$i=0 \to vector.size$} {
		\If{$vector[i] \geq 0$} {
			$fingerprint[i]$ = true
		}
		\Else {
			$fingerprint[i]$ = false
		}
	}
	emit(KEY($sequenceID$),VALUE($fingerprint$))
\end{algorithm}

\section{Implementation in MapReduce}
\label{sec:MapReduce}

As has been shown in Section~\ref{sec:LSH}, LSH is essentially a signature-producing algorithm, or a sketching algorithm that produces a compact representation of a whole data sequence so that measurements on the original data sequence can be estimated by efficient computations on the compact representation, which in this case is the signature of the corresponding sequence. The implementation of ScalLoPS is to be carried out in two phases, the first being the signature generation, that produces the compact representations of the sequences, while the second being the signature processing, that performs the efficient computation on the compact representations. Both algorithms are implemented in MapReduce paradigm in order to achieve scalability in high performance.

The MapReduce programming model consists of two functions called Map and Reduce. Data are distributed in Key and Value pairs on which the Map function computes a different set of intermediate Key and Value pairs. These pairs are then reduced appropriately by the Reduce function which performs computation on the lists of intermediate Values that have the same intermediate Key. The resulting set of Key and Value pairs from the reducers is the final output.

Therefore, the complete framework of the protein sequence searching in ScalLoPS is as following. In the beginning, the MapReduce Signature Generator is utilised to generate the complete sets of signatures for each of the protein sequence sets. Then, when a query set of protein sequences for similar search is available, the MapReduce Signature Generator firstly generates a complete set of signatures for the set. Then, the MapReduce Signature Processor compares the signatures of both the query set and the reference set in order to produce pairs of sequences that are similar. The following subsections explain both the Signature Generator and the Signature Processor in more detail.

\subsection{MapReduce Signature Generator}

Since each signature is generated out of its own sequence and is not depending on other sequences, then scalability  can be achieved effectively by processing each sequence individually. Implementing in MapReduce fashion, this signature generating function can therefore be efficiently implemented as the map function, while the reduce function is not required. In essence, the reduce function becomes an identity function (a pass-through), that passes the input Key and input Value into output Key and output Value without any processing. As shown in Algorithm~\ref{alg:SimhashSigGen}, the input Key to each map worker is the sequence ID, while the input Value is the sequence string. Then, the signature generation is performed by the LSH algorithm mentioned in Section~\ref{sec:LSH}, and the resulting signature will be the output Value, and the output Key is the same sequence ID that serves as the input Key.

\begin{figure}[t]
   \centering
   \subfigure[Signature Generator]{
     \includegraphics[width=0.45\textwidth]{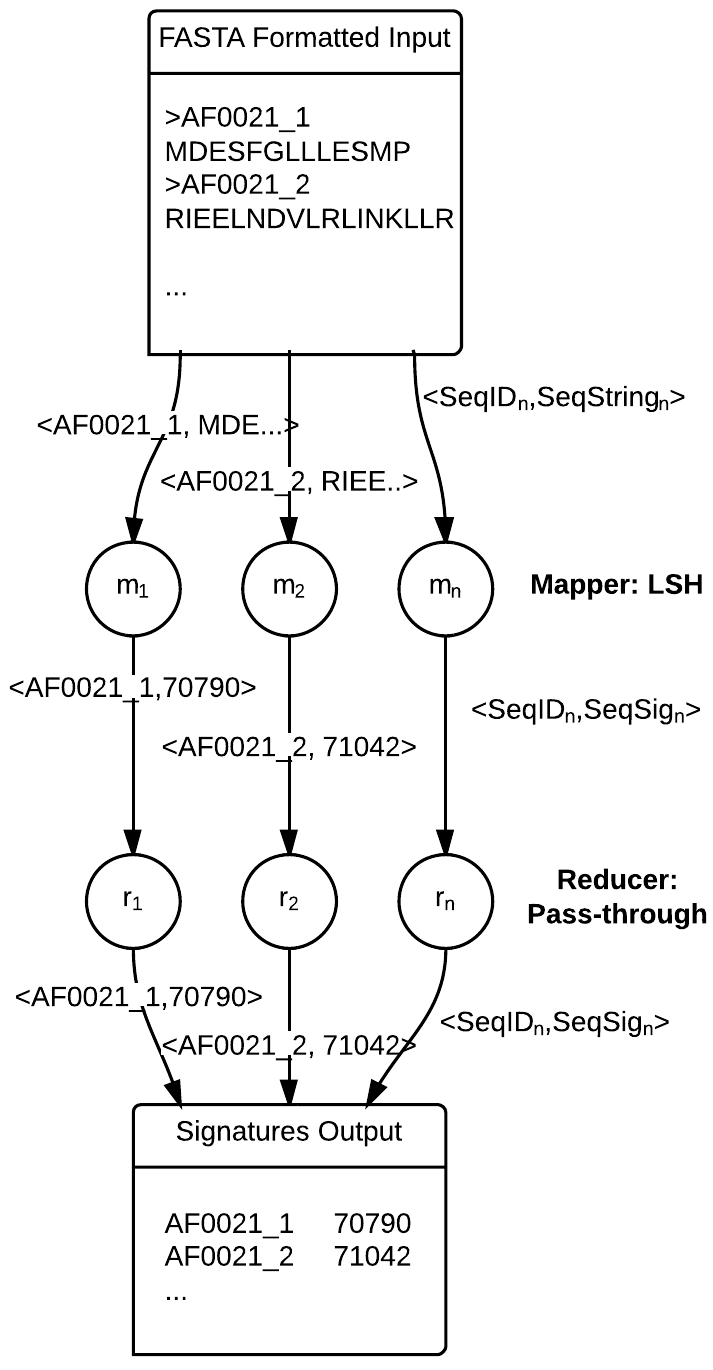}
     \label{fig:SigGenMR}}
   \subfigure[Signature Processor]{
       \includegraphics[width=0.45\textwidth]{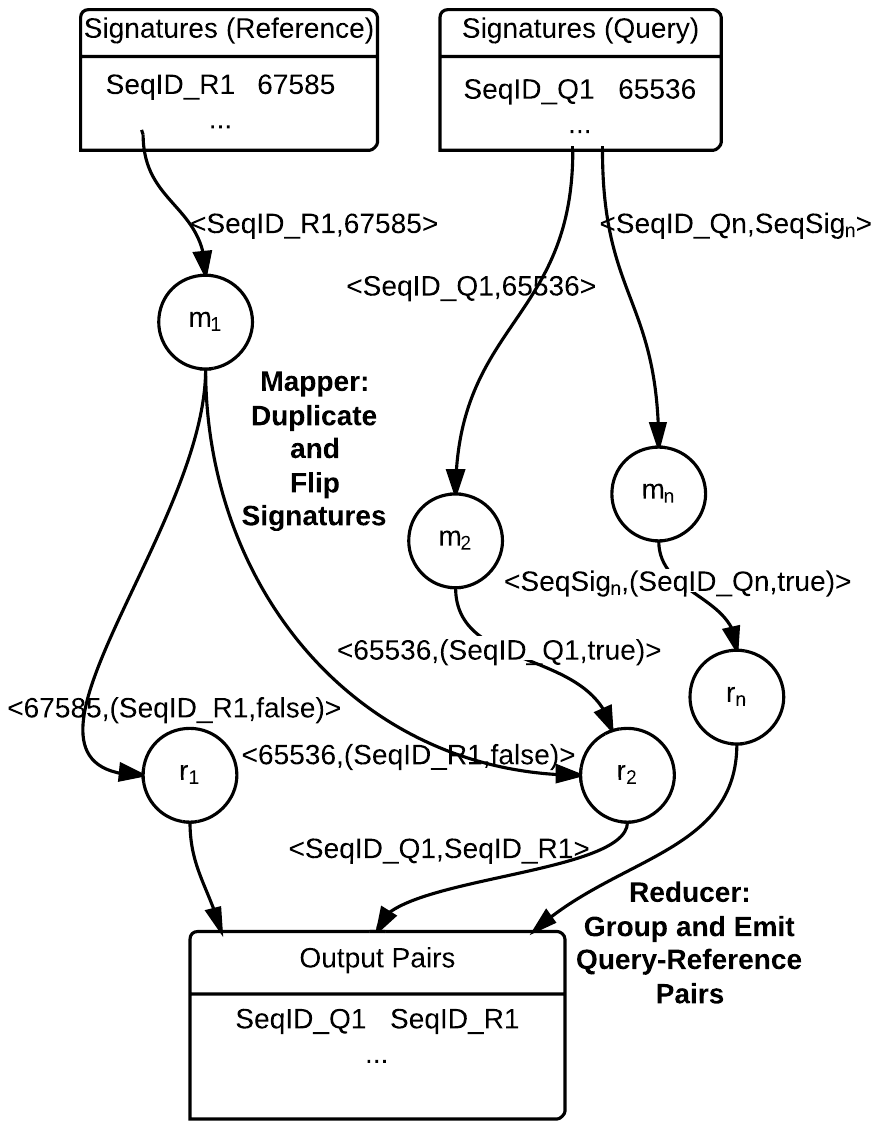}
       \label{fig:SigProcMR}
     }
     \caption{The MapReduce workflow in ScalLoPS}  
\end{figure}

In the example of Figure~\ref{fig:SigGenMR}, the mapper \texttt{m$_{1}$} receives the first protein sequence in the input file, \texttt{MDESFGLLLESMO} as its input Value with sequence ID \texttt{AF0021\_1} as its input Key, and then emits the same ID \texttt{AF0021\_1} as its output Key, and the signature \texttt{70790} as produced by the Algorithm \ref{alg:SimhashSigGen}, is emitted as its output Value. This sequence ID and signature will then pass through the identity reducer \texttt{m$_{1}$} and be written in the output file. Similarly, at the same time the mapper \texttt{m$_{2}$} receives the next protein sequence \texttt{RIEELNDVLRLINKLLR} in the list with sequence ID \texttt{AF0021\_2} and emits its sequence ID along with its signature \texttt{71042} to be passed through by the identity reducer \texttt{m$_{1}$} and be written in the output file.

\subsection{Signature Processor}
The signature processing is carried out to compare the signatures of two different sets of sequences, usually designated as the query set and the reference set respectively, where the reference set is a set of the sequences that is queried against the sequences in the query set. Therefore, pairs of query sequence and reference sequence which are similar, having their signature Hamming distance less than a set threshold, will be emitted as the final result.

To achieve scalability, comparison will be carried out by exploiting the MapReduce programming paradigm of grouping together all objects with the same Key in the reduce phase.

Here the map function would take sequence ID and the corresponding sequence signature as its input Key-Value pair. Then, as shown in Algorithm \ref{alg:SigProcMapFunction}, for every sequence, the map function will emit the sequence signature as the output Key and a special object wrapping the sequence ID and a boolean value (that is to indicate if the sequence is a query or a reference) as the output Value. However, for every reference sequence only, the map function will as well call a \texttt{flip()} function that returns a list of integers whose bits are at most $t$ Hamming distance away from the signature. The function performs by looping through all possible bit combinations given the Hamming distance. Then, for each integer returned by the function, the mapper will emit the integer as the output Key, and the same object that contains the sequence ID and a false boolean value as the output Value. 

\begin{algorithm} [t]
\SetAlgoLined
 	\KwData{KEY($seqID$),VALUE($seqSig$)}
	\If{$fileName$='query'}{
		emit(KEY($seqSig$), VALUE($seqID$,true))
	}
	\Else{
		emit(KEY($seqSig$), VALUE($seqID$,false))\\
		$flippedSigs$=flip($seqSig$)\\
		\For{$flippedSig \in flippedSigs$}{
			emit(KEY($flipped$),VALUE($seqID$,false))
		}
	}
 	
\caption{Signature Processor Map Function}
\label{alg:SigProcMapFunction}
\end{algorithm}

Having received from the mapper a signature or its flip as the input Key, and as the input Value a list that contains a query sequence ID, or possibly more query sequence IDs with the same signature, along with all the reference IDs which individual signature is $t$ Hamming distance away from the query signature, as shown in Algorithm \ref{alg:SigProcReduceFunction}, the reduce function then loop through the list of reference IDs. Then, in each loop the function emits the query ID as the output key and the reference ID as the output value. If there is more than one query ID, then the reduce function will repeat this process for all the rest of the query IDs. Therefore, no duplication of pair is guaranteed.

\begin{algorithm} [t]
\SetAlgoLined
 	\KwData{KEY($sig$), VALUE(LIST[$seqID$, $isQuery$])}
	$queries \gets$ \{\}\\
	$references \gets$ \{\}\\
	\For{$value \in ValueList$}{
		\If{$value.isQuery$}{
			$queries$.add($value.seqID$)
		}
		\Else{
			$references$.add($value.seqID$)
		}
	}
	\For{$query \in queries$}{
		\For{$reference \in references$}{
			emit(KEY($query$),VALUE($reference$))
		}
	}
 \caption{Signature Processor Reduce Function}
\label{alg:SigProcReduceFunction}
\end{algorithm}

An example in Figure~\ref{fig:SigProcMR} below shows two sequences, one from the reference set, designated as \texttt{SeqID\_R1} with \texttt{67585} as its signature, and the other is from the query set, designated as \texttt{SeqID\_Q1} with \texttt{65536}. Then \texttt{SeqID\_R1} is read by mapper task \texttt{m$_{1}$}, which then emits the signature itself as long as many flips of the signature, one of them being \texttt{65536}, which is also the signature of \texttt{SeqID\_Q1}. Therefore, at the reduce phase, this flipped signature of \texttt{SeqID\_R1} is received by the same reducer (\texttt{r$_{2}$}) as the signature of \texttt{SeqID\_Q1}. The reducer then emits \texttt{SeqID\_Q1} and \texttt{SeqID\_R1}, recognising both as a similar pair.

\section{Experimental Evaluation}
\label{sec:Evaluation}
We carried out the experiments to measure the quality, performance, and scalability of ScalLoPS. Quality is evaluated by measuring the changes in the percentage of identitiesin the alignments over varying parameters. Performance, on the other hand, is evaluated by taking into account the running time of both the signature processor and the signature generator compared to that of BLAST and RAPSearch. Finally, scalability is evaluated by measuring the changes in running time over different number of processing nodes.

The parameters for our experiments are the shingle length $k$, that is the length of the substrings of the protein sequences that produce the neighbouring words on both BLAST and ScalLoPS, and the neighbouring word lower bound threshold $T$. These parameters govern the tradeoff between sensitivity and performance. Higher $T$ will result in a lower processing time but less sensitivity and vice versa, while higher $k$ will result in higher processing time with higher sensitivity and vice versa. By default, BLAST uses $k$=3 and $T$=11 for protein sequences in order to produce sensitive results with reasonable performance. Furthermore, in ScalLoPS, there is another parameter called the Hamming distance threshold, $d$, that is also the trade off of performance and sensitivity. Higher $d$ will result in a higher processing time and higher sensitivity, and vice versa.

The next subsection describes the input query and reference datasets that we have used for the experiments. Then, we discuss the results for each experiment in detail and present the inferences gathered from these.

\subsection{Datasets}

There are three query sets used in our experiment listed in Table~\ref{tbl:Querysets}. One of the sets is called \texttt{NC\_000913.faa}, which contains curated protein sequences from the well-studied \emph{Escherichia coli} K12 MG1655. The second set called \texttt{227\_01\_prot}, which came from a sample taken from the deepest layer of stratified Ace Lake in East Antarctica, that are mostly uncharacterised as their relatives could not be found in established databases~\cite{227}.

The final query set is obtained from The Sorcerer II Global Ocean Sampling (GOS) expedition that was carried out from northwest Atlantic through eastern tropical Pacific~\cite{GOS}. This DNA sequence dataset is available for download from the website of Community Cyberinfrastructure for Advanced Microbial Ecology Research and Analysis (CAMERA)~\cite{CAMERA}. We then applied a tool known as MetaGene~\cite{Metagene} on the raw DNA sequence reads of the sample taken from this expedition in order to predict the whole range of the genes coded in the DNA. Finally, we translated the DNA sequences into a set of protein sequences that we termed \texttt{allgos}.


\begin{table}[t]
\centering
\caption{Query sets used for the experiment}
\label{tbl:Querysets}
\begin{tabular} { | c | c | c | c |}
\hline
Query set & No. sequences & Size (MB) & Average length\\
\hline
\texttt{NC\_000913.faa} & 4,146 & 1.73 & 316\\
\texttt{227\_01\_prot} & 547,169 & 87.42 &  80.93\\
\texttt{allgos} & 120,723,333 & 13,750.2 & 24.12\\
\hline
\end{tabular}
\end{table}

\begin{table}[t]
\centering
\caption{Reference datasets used for the experiment}
\label{tbl:Datasets}
\begin{tabular} {| c | c | c | c | }
\hline
Dataset & No. sequences & Size (MB) & Average length\\
\hline
\texttt{myva} & 192,987 & 58.805 & 305\\
\texttt{swissprot} & 454,401 & 232.28 & 372.85\\
\texttt{nr} & 23,074,873 & 13,680.9 & 343.38\\
\hline
\end{tabular}
\end{table}

The reference sets include \texttt{myva}, a set of FASTA-formatted protein sequences containing all the proteins from 66 complete genomes classified phylogenetically into Clusters of Orthologous Groups (COG), \texttt{swissprot}, a database of high quality annotated and non-redundant protein sequences maintained by the UniProt consortium, and \texttt{nr}, a database of non-redundant protein sequences composite of SwissProt, SwissProt Updates, Protein Information Resource (PIR), and Protein Data Bank (PDB). The number of sequences, size, and  average length of each of these reference sets are shown in Table~\ref{tbl:Datasets}.

\subsection{Quality}
\label{sec:quality}

In this experiment, we demonstrate the effects of varying different parameters, such as shingle length $k$, neighbouring word score threshold $T$, and Hamming Distance $d$ on the quality of the results returned by ScalLoPS for different query sets and reference sets, and compare it to the results returned by BLAST for the same inputs. The experiments in this section were carried out on a private cloud in the School of Computer Science and Engineering, UNSW running OpenStack cloud management system. We executed the entire MapReduce workflow on five instances, each of which ran Ubuntu 12.04.1 64-bit with a single virtual 2.4 GHz CPU core, 2 GB of RAM, and 200 GB of volume storage (used for HDFS). Additionally, the data was stored in HDFS with replication factor of 2.


We use the measure of \emph{percent identity (PID)}, which is the number of the same residues at the same positions expressed as a percentage of the total number of the residues in an alignment between a query and a reference. The PID is a measure of the quality of the alignment produced by a tool, and a higher overall PID indicates better results.  However, there is a huge variation in the PIDs obtained from searching a query set against a reference set. Therefore, we present the PIDs of the query results as box plots. Box plots are a convenient method to depict the distribution of a set of data points in terms of quartiles. The statistics shown are minimum (Q0), lower quartile (Q1=25th percentile), median (Q2=50th percentile), upper quartile (Q3=75th percentile), and the maximum (Q4). These plots enable the comparison of not only the range but also the skew of PIDs produced by each method for the same query and reference sets.

\begin{figure}[t]
\centering
\subfigure[Effect of varying Hamming distance threshold on the percentage of identities of all pairs] {
\includegraphics[width=0.35\textwidth]{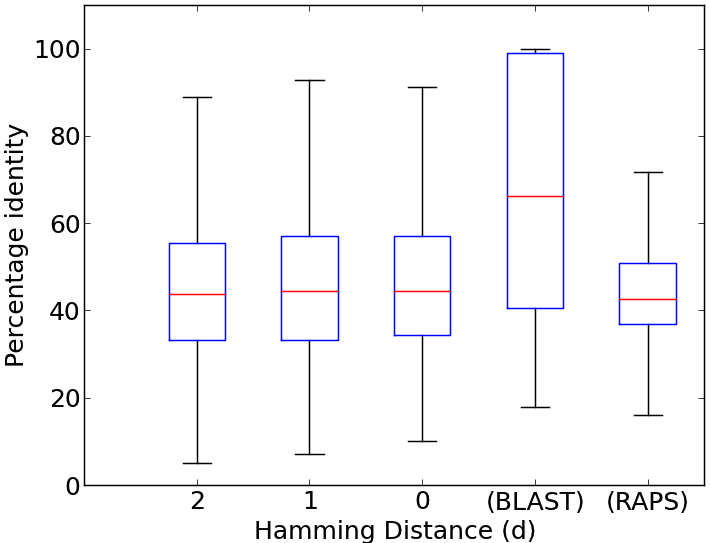}
\label{fig:AvsBperID}
}
\subfigure[Effect of varying Hamming distance threshold on the resulting number of pairs] {
\includegraphics[width=0.35\textwidth]{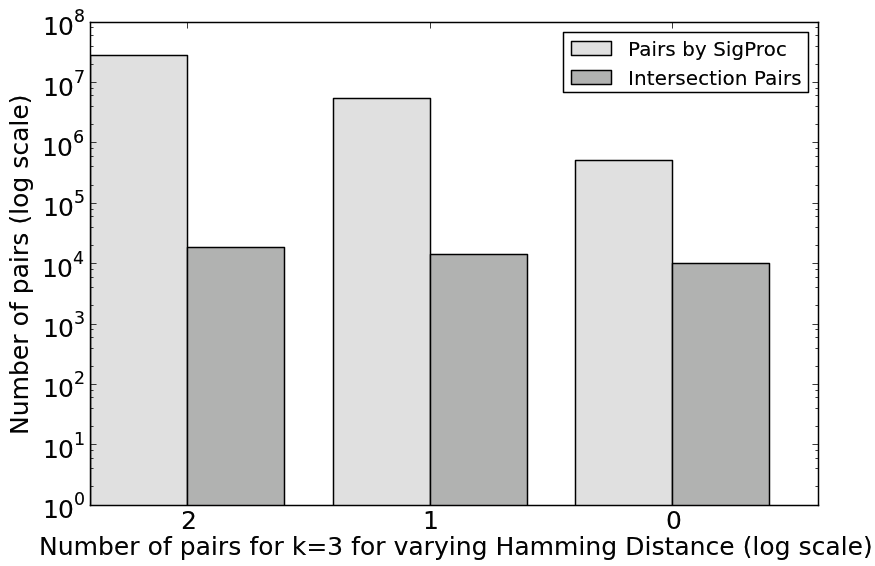}
\label{fig:NumPairsvarHD}
}
\subfigure[Effect of varying Hamming distance threshold on the percentage identities of intersection pairs only] {
\includegraphics[width=0.35\textwidth]{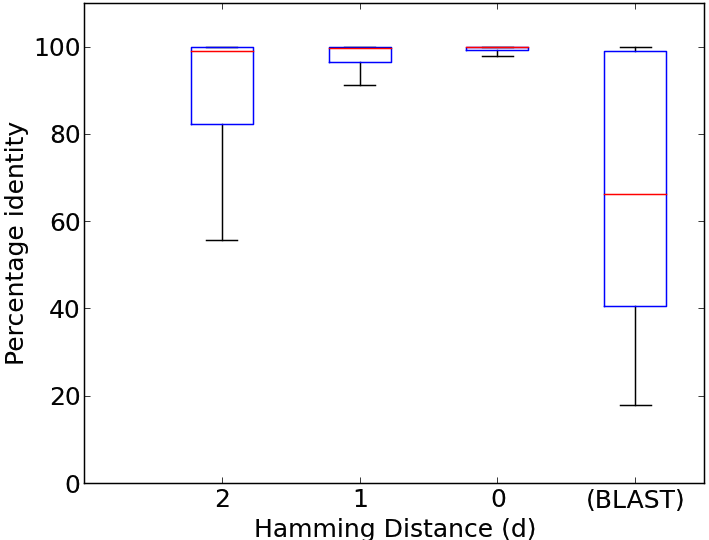}
\label{fig:AnBperID}
}
\caption{Effect of varying Hamming distance threshold on percentage of identities and number of pairs}
\label{fig:exp-quality-varHD}
\end{figure}

We carried out our experiments first on the smallest combination of query and reference sets, \texttt{NC\_000913.faa} versus \texttt{myva}. Figure~\ref{fig:AvsBperID} shows the box plots of the PIDs of the query-reference pairs produced by the ScalLoPS Signature Processor, under Hamming distance thresholds $d = \{0, 1, 2\}$. We also executed the same search using BLAST and RAPSearch. On an average, ScalLoPS produced larger sets of results (around 1-2 orders of magnitude higher) than BLAST or RAPSearch. The median PID for BLAST is around 65\% while that for the others, the same measure is around 45\%. Also, 75\% of BLAST results (between Q4 and Q1) have above 45\% PID, indicating higher quality of results overall. 

The expected success of BLAST can be attributed to the statistical measures used in BLAST to aggressively reduce the number of alignments with low similarities. ScalLoPS presents unfiltered results and therefore, has a much more larger range of PIDs. To compare the results of BLAST and ScalLoPS evenly, we isolated the query-reference pairs that were found in both ScalLoPS and BLAST results, which we term here as \emph{intersection pairs}. Looking at these pairs enables us to discover the proportion of ScalLoPS that was in common with BLAST and the percentage identity of that portion. A high value for both these measures indicates that ScalLoPS captures the pairs with the best alignment scores in the BLAST results.

Figure~\ref{fig:NumPairsvarHD} and Figure~\ref{fig:AnBperID} show the intersection pairs as a proportion of the total ScalLoPS results and the box plots for the PIDs of the pairs respectively, for different values of Hamming distance $d$. At $d = 2$, only a small proportion of ScalLoPS results were in common with BLAST with the majority of the intersection pairs (between Q1 and Q4), falling between 80-100\% identity. As $d$ is reduced, the proportion increases and the overall size of the results is reduced. 

These results are consistent with the approximation of cosine distance by the random hyperplane techniques. The positions of each of both the feature vectors of the pair relative to the hyperplanes form the signatures, whose Hamming distance is almost equivalent to the cosine distance of both the pair vectors. Hence, the lower the Hamming distance, the lower the number of possible false positives, and hence the overall size of results is lower. 

It can be seen that lowering the Hamming distance tightens the PID range of the intersection pairs. For $d=0$, the intersection pairs have a very high median PID (between 95-100\%). Overall, Figure~\ref{fig:NumPairsvarHD} also shows that lower Hamming distance actually results in lower number of pairs produced by Signature Processor, while maintaining almost the same number of intersection pairs. This means that the results with the highest quality and relevance are obtained at $d=0$. Therefore, the subsequent experiments are conducted with $d=0$.


\begin{figure}[t]
\centering
\subfigure[Effect of varying neighbourhood score threshold on the percentage identities of intersection pairs only] {
\includegraphics[width=0.4\textwidth]{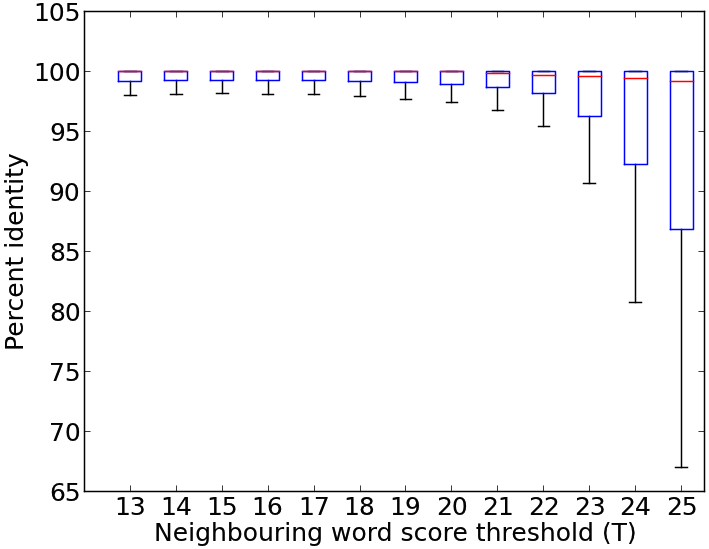}
\label{fig:AnBvarT}
}
\subfigure[Effect of varying neighbourhood score threshold on the resulting number of pairs] {
\includegraphics[width=0.4\textwidth]{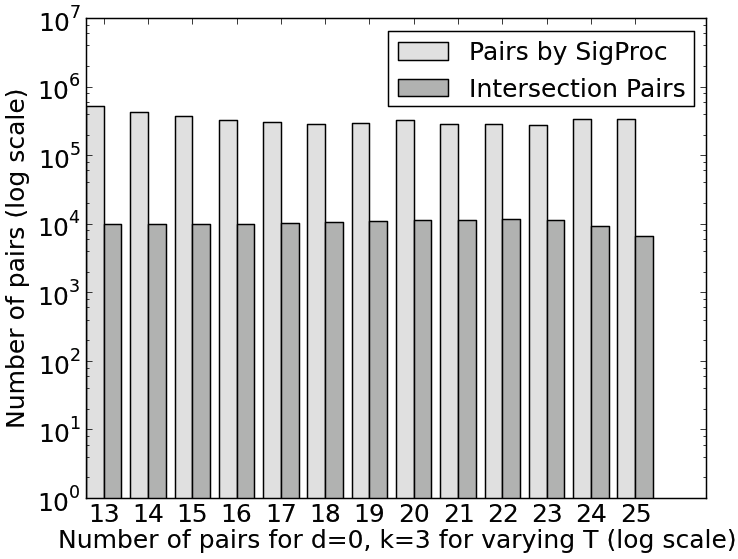}
\label{fig:NumPairsvarT}
}
\caption{Effect of varying neighbourhood score threshold on percentage of identities and number of pairs}
\label{fig:exp-quality-varT}
\end{figure}

Figure~\ref{fig:AnBvarT} and Figure~\ref{fig:NumPairsvarT} show the effect of varying $T$ on the PIDs of the intersection pairs. It can be perceived that the median PIDs are high and stable from $T=13$ to $T=20$, and start to decrease from $T=21$, while the number of similar pairs produced by Signature Processing decreases and the number of intersection pairs increases as $T$ increases from 13 to 18. Then from $T=19$ onwards, the number fluctuates.

This may be explained by the utilisation of neighbouring words as features. As $T$ increases, less neighbouring words are generated, and therefore less features are being utilised. Therefore, the distortion caused by neighbouring words is minimised as $T$ increases. However, at very high $T$, no neighbouring words are generated at all, and the signatures of such sequences converge to all zeroes. Although the Signature Processing is designed to process only the sequences with non-zero signatures, fluctuations can still be perceived at high $T$, which may be caused by the lack of low value feature weights.

Both Figure~\ref{fig:AnBvarT} and Figure~\ref{fig:NumPairsvarT} mentioned above indicate that the neighbouring word score threshold $T=22$ results in the lowest number of possible false positives with reasonable median PIDs, and as such the subsequent experiments are conducted with this threshold value.

\begin{figure}[t]
\centering
\subfigure[Effect of varying shingle length on the percentage identities of intersection pairs only] {
\includegraphics[width=0.4\textwidth]{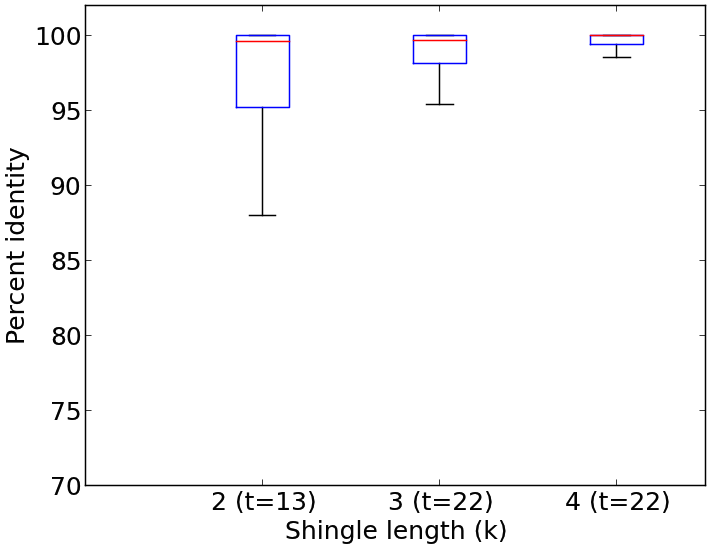}
\label{fig:AnBvark}
}
\subfigure[Effect of varying shingle length on the resulting number of pairs] {
\includegraphics[width=0.4\textwidth]{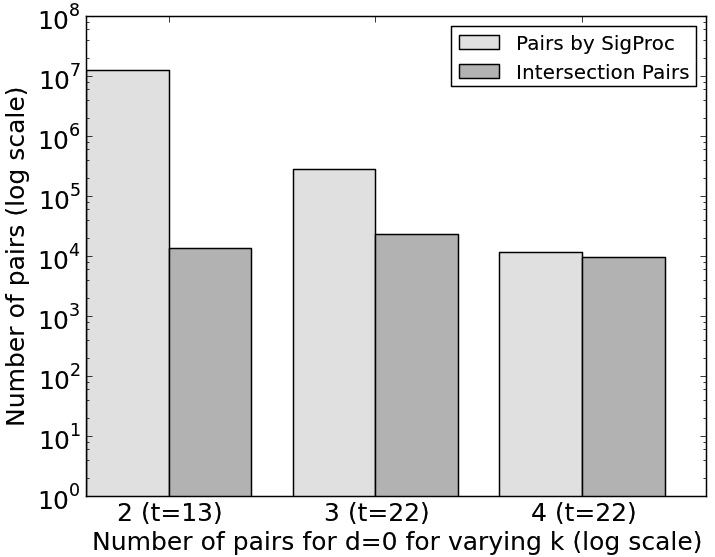}
\label{fig:NumPairsvark}
}
\caption{Effect of varying shingle length on percentage of identities and number of pairs}
\label{fig:exp-quality-varK}
\end{figure}

Figure~\ref{fig:AnBvark} and Figure~\ref{fig:NumPairsvark} show the effect of varying $k$ on the PIDs and the number of similar pairs generated. Specifically for $k=2$, $T$ is chosen to be a lower value ($T=13$) since at $T=22$ there is no neighbourhood word being generated at all, including the parent words, and therefore the signatures converge to all zeroes, and this in turn will result in all sequences being treated as undesirably equal. 

As $k$ increases from 2 to 4, the median PIDs increase, and consistently, the number of similar pairs generated by the Signature Processor reduces significantly to be almost equivalent with the number of intersection pairs while the number of intersection pairs decreases slightly. The increase of $k$ therefore reduces the total number of features that can be generated from a sequence. Both Figure~\ref{fig:AnBvark} and Figure~\ref{fig:NumPairsvark} suggest that $k=4$ produces more desirable sequence pairs with higher median PIDs and lower number of possible false positives.

\begin{figure}[t]
\centering
\subfigure[Applying parameters k=4, t=22, and d=0 to different combinations of sets] {
\includegraphics[width=0.4\textwidth]{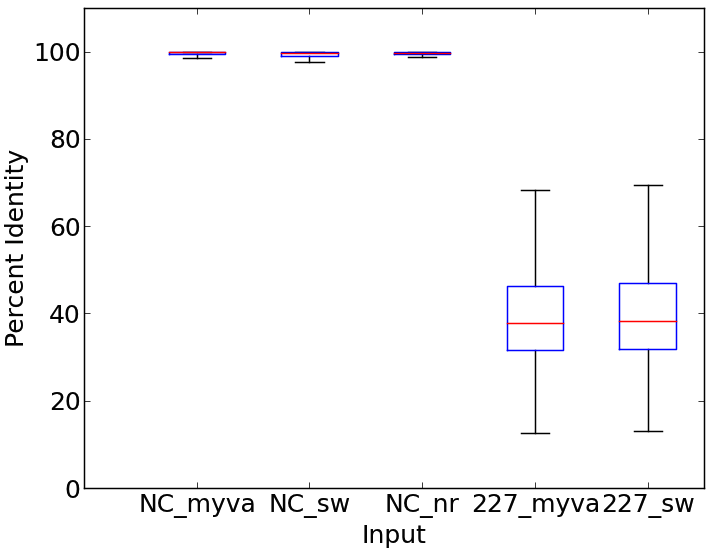}
\label{fig:AnBvarSet}
}
\caption{Effect of varying different parameters on percentage of identities and number of pairs}
\label{fig:exp-quality-varSet}
\end{figure}

Figure~\ref{fig:AnBvarSet} shows the application of these selected parameter values of $d$, $T$, and $k$ for different combinations of datasets and the resulting PIDs of the intersection pairs. The figure suggests that the method produces the desired high median PIDs with \texttt{NC\_000913.faa} while it produces significantly lower median PIDs for \texttt{227\_01\_prot}. This may be explained by the average sequence length of the datasets. The average sequence lengths of each of the sets \texttt{NC\_000913.faa}, \texttt{myva}, \texttt{swissprot}, and \texttt{nr} are all in the range of  300-400, while the average length of the sequences in \texttt{227\_01\_prot} is below 100. The high difference in sequence average lengths would reverses the sign of the dot product of the sequences being compared, which in turn affects the signatures generated, and therefore increases the number of false positives.

As an example, suppose that the feature vector for sequence \texttt{$s_1$} is \texttt{(\{A,B,C\}, \{B,C,D\}, \{C,D,E\})}, and the feature vector for sequence \texttt{$s_2$} is \texttt{(\{B,C,D\}, \{C,D,E\}, \{A,B,C\}, \{D,E,F\},\{E,F,G\})}, a shortened version of \texttt{$s_2$} designated as \texttt{$s_3$} which feature vector is \texttt{(\{B,C,D\}, \{C,D,E\}, \{A,B,C\}}, and the weight of each feature is one. Suppose also that the hash value of \texttt{\{A,B,C\}} is \texttt{\{1,0,0,1\}}, \texttt{\{B,C,D\}} is \texttt{\{0,1,1,1\}}, \texttt{\{C,D,E\}} is \texttt{\{1,1,0,1\}}, \texttt{\{D,E,F\}} is \texttt{\{0,1,0,1\}}, and \texttt{\{E,F,G\}} is \texttt{\{0,0,1,1\}}. In this example, both \texttt{$s_1$} and \texttt{$s_3$} produces the same signature \texttt{\{1,1,0,1\}} and therefore correctly predicted as similar, while \texttt{$s_2$} produces the signature \texttt{\{0,1,0,1\}} which is not included in the list of similar pairs when using $d$=0. Therefore the number of false negatives increases, and this in turn decreases the median PID.

\subsection{Performance}
In these experiments, we demonstrate that ScalLoPS generates the query-reference pairs with a reasonable time compared to that of BLAST for small sets, and with a significantly lower processing time for large sets. Our experiments were carried out on the same private cloud infrastructure as presented in the previous Quality experiments. The experiments on BLAST are carried out with default parameters of $k$ and $T$, while the experiments of ScalLoPS were carried out with $k=3$ and $t=13$, and $d=2$ for the ScalLoPS Signature Processor.

The experiments on BLAST and RAPSearch are carried out by firstly preparing the database using the reference sequences with \texttt{makeblastdb} for BLAST and \texttt{prerapsearch} for RAPSearch respectively. Then, the database is copied and distributed to all the processing nodes. After that, the input query set is split into the number of processing nodes and each of the splits is distributed to a different node. Finally, BLAST or RAPSearch is performed in each of the processing nodes on the split query set and the database, and the time required by the node with the longest processing time was reported in Table~\ref{tbl:PerformanceImprovement}.

The time required by BLAST and RAPSearch as reported in Table~\ref{tbl:PerformanceImprovement} below excludes the time taken for the database preparation and query splitting, and the distribution of both the database and the query splits to all the processing nodes. The time required by ScalLoPS is also reported in Table~\ref{tbl:PerformanceImprovement}. As the signature generation for the reference sequences is analogous to the database preparation for the other two tools and has to be done only once, the time required by ScalLoPS is reported as the sum of the time required for generating the signatures for the query sequences and the time required to produce the query-reference pairs.

\begin{table}[!ht]
\centering
\caption{The time taken for ScalLoPS, BLAST, and RAPSearch to produce similar protein pairs (in minutes)}
\label{tbl:PerformanceImprovement}
\begin{tabular} { | c | c | c | c | c |}
\hline
Query vs Reference & ScalLoPS & BLAST & RAPSearch \\
\hline
\texttt{NC\_000913.faa} vs \texttt{myva} & 19.77 & 13.09 & 1.14\\
\texttt{NC\_000913.faa} vs \texttt{swissprot} & 61.52 & 35.11 & 4.24\\
\texttt{227\_01\_prot} vs \texttt{myva} & 100.46 & 372.12 & 10.58\\
\texttt{227\_01\_prot} vs \texttt{swissprot} & 291.55 & 1004.1 & 27.9\\
\hline
\end{tabular}
\end{table}

As shown in Table~\ref{tbl:PerformanceImprovement}, for small datasets, ScalLoPS takes a longer time for generating results than either BLAST or RAPSearch. However, as the datasets become larger (such as \texttt{227\_01\_prot}), ScalLoPS produces better performance than BLAST, though still significantly slower than RAPSearch. ScalLoPs is therefore better suited for studying metagenomics datasets, such as those produced by Global Ocean Sampling expedition, which are very large compared to datasets produced by earlier genomic studies.

It should also be noted that all the data formatting and distributions of BLAST and RAPSearch have to be carried out manually, since these were not designed for scaling across multiple processing nodes. Moreover, results produced by distributing BLAST and RAPSearch in this manner are different from single machine executions since the calculations of measures such as the expect value (e-value) and bit-score depend on retaining the entire query set on a single machine. However, these problems with distributed BLAST execution have been largely solved in mpiBLAST. In contrast, ScalLoPS leverages the data management facilities provided by the Hadoop framework that take care of data distribution and replication. 

The difference shown in this performance measurement is also a result of the difference in computational complexity of BLAST and ScalLoPS. According to Altschul, et. al~\cite{BLAST}, the computational complexity of BLAST is approximately $O(aW + bN + cN W/20k)$ where $W$ is the number of neighbouring words generated, N is the number of residues (letters) in the datasets, $k$ is the shingle length, and $a$, $b$, and $c$ are constants. In contrast, the computational complexity of our ScalLoPS-based signature generator is $O(W)$ and the complexity of our signature processor is $O(Y)$ where $Y$ is the number of the original signatures and the flipped signatures in the datasets. Since reference signatures are reusable for different queries, then the computational complexity for the whole query processing is $O(W)+O(Y)$.

\subsection{Scalability}

\begin{figure}[t]
\centering
\subfigure[Effect of varying the number of cores on signature generating time] {
\includegraphics[width=0.42\textwidth]{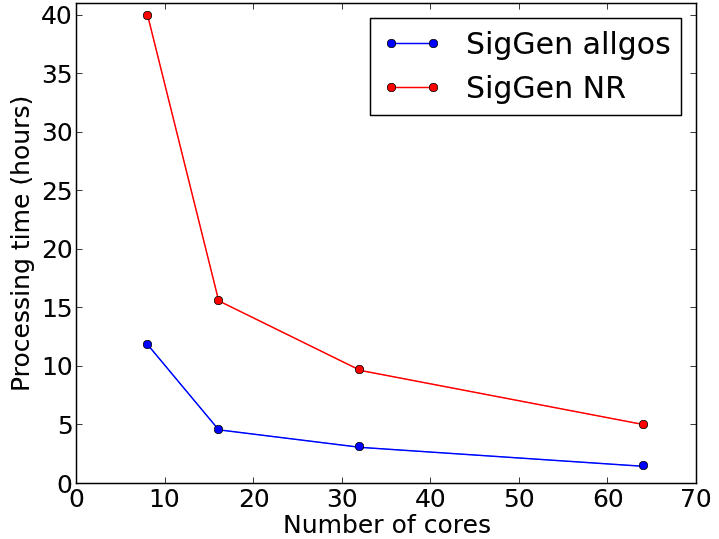}
\label{fig:SigGen-scal}
}
\subfigure[Effect of varying the number of cores on signature processor time] {
\includegraphics[width=0.42\textwidth]{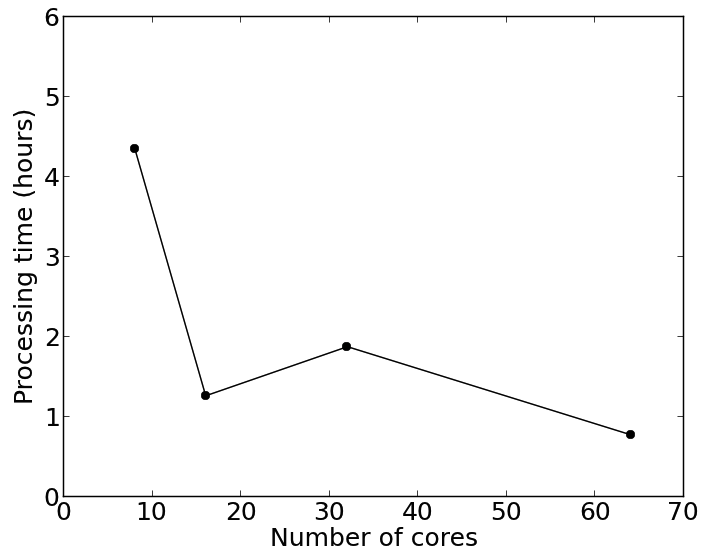}
\label{fig:SigProc-scal}
}
\caption{Effect of varying number of cores on ScalLoPS}
\label{fig:exp-scalability}
\end{figure}

In this section, we demonstrate the effects of varying the number of processing nodes on the processing time on both the Signature Generator and the Signature Processor. The input datasets for the experiments were \texttt{allgos}, the biggest query set and \texttt{nr}, the largest reference set. Executing a BLAST search for \texttt{allgos} against \texttt{nr} on a single dual-core VM resulted in a wall-clock time of approximately 14 days. Therefore, this is a computationally-intensive workflow that requires distributed resources. 

Due to lack of resources in the private cloud infrastructure, this set of experiments were carried using Amazon Elastic MapReduce (EMR). The EMR service automatically creates and deploys Hadoop MapReduce processes on resources sourced from the Amazon Elastic Cloud Compute (EC2) IaaS service. The input data for EMR is stored by the user in Amazon S3 (Simple Storage Service), which is also the destination for output data. However, the user has to supply the actual MapReduce workflow that is to be executed by the EMR service.

We created a combined distribution of ScalLoPS (Signature Generator plus Signature Processor) that was then deployed using EMR. ScalLoPS was executed with the parameter values $k$=4, $T$=22, and $d$=0 that were demonstrated as leading to the best quality results in Section~\ref{sec:quality}. We utilised EC2 Extra Large instances~\footnote{Amazon EC2 instances - http://aws.amazon.com/elasticmapreduce/\#instance} that had 4 cores of 64-bit virtual CPU (equivalent to 8 Elastic Compute Units, each is equivalent to 1.0-1.2 GHz Opteron or Xeon processor), 15 GB memory, and 840 GB instance storage. The experiments were carried out by doubling the number of instances, starting with 2 instances up to a maximum of 8. Figures~\ref{fig:SigGen-scal} and \ref{fig:SigProc-scal} show the wall-clock times taken for the Signature Generator and the Signature Processor with increasing number of processing cores.

Figure~\ref{fig:SigGen-scal} shows an almost inverse-exponential decrease in signature generating time for each of both \texttt{allgos} and \texttt{nr} as the number of processing cores increases. This decrease confirms that ScalLoPS is scalable to the number of processing nodes. As mentioned in Section~\ref{sec:MapReduce}, the Signature Generator MapReduce contains Map function only, which receives one sequence and produces a signature out of the sequence. Therefore, the MapReduce program is able to distribute the input dataset into as many number of map workers available at a time. Since the number of map workers corresponds directly to the number of processing cores, the time taken decreases as this number increases. 


Figure~\ref{fig:SigProc-scal} also shows an almost inverse-exponential decrease in signature processing time for \texttt{allgos} vs \texttt{myva} as the number of processing cores increases with a fluctuation at 32 cores. Further examinations from the log file have indicated that this is caused by several task workers reached a set timeout, failing to report back to the master worker such that the tasks had to be restarted. This situation is therefore, an artifact of a particular execution.


\section{Conclusion and Future Works}
\label{sec:FutureWorks}
To summarize, this paper introduces a new tool called ScalLoPS that improves the scalability of protein homology searching by implementing algorithms that leverage Locality-Sensitive Hashing and the MapReduce paradigm. This tool is motivated by the data analysis challenges in the emerging field of metagenomics and the ability to use cloud computing to solve them. 

ScalLoPS generates signatures for each sequence in the query and reference sets, and then measures the Hamming distance between them. Through extensive experiments using datasets derived from actual metagnomic studies, we have shown the quality, performance and scalability characteristics of ScalLoPS against BLAST, the standard bioinformatics too for sequence searching and analysis. 

Our first results shows the effect of parameters such as shingle length $k$, threshold $T$, and Hamming distance $d$ on the quality of the result produced by ScalLoPS. Higher $k$ and $T$, and lower $d$ will result in less number of pairs produced but with higher percent identity.  However, query sets with smaller average sequence lengths will produce less percentage identity than those with longer sequences. We also show that, with the right parameter values, ScalLoPS captures a high proportion of the query-reference pairs with high alignment identity, produced by BLAST. 

We also demonstrate that ScalLoPS has better performance than BLAST for large query sets when distributed across multiple nodes. The scalability experiments executed on actual cloud resources demonstrate that ScalLoPS is able to utilise distributed resources effectively. The scalability of ScalLoPS is seamless, since it leverages the MapReduce framework and HDFS that provide data storage and distribution for any number of nodes. Therefore, we conclude that ScalLoPS satisfies the requirements of a scale-agnostic tool that is able to take advantage of cloud resources to deliver high-performance protein sequence searching.

Efforts are on to reduce the number of false positives produced by ScalLoPS further in order to achieve higher quality results. This can be achieved by actually running an alignment algorithm and filtering out pairs with lower quality as measured by the alignment length and the alignment percentage of identity. We would like to implement a distributed method of calculating the expect value and bit-score so that ScalLoPS can be used as a substitute for BLAST. We would also like to improve the performance of ScalLoPS by applying the reduced alphabet method employed by RAPSearch.





\bibliographystyle{plain}
\bibliography{ConfPaperBib}

\end{document}